\documentclass[aps,prl,amsmath,amssymb,preprintnumbers,twocolumn,showpacs,floatfix,groupedaddress]{revtex4}
\usepackage{epsfig,graphics}
\usepackage{graphicx}
\usepackage{dcolumn}
\usepackage{bm}
\usepackage{amsmath}
\usepackage[usenames]{color}
\usepackage{ulem} 

\newcommand \vk {\vec{k}}

\newcommand \vp {\vec{p}}

\newcommand \hf {\frac{1}{2}}

\newcommand \prt {\partial}

\newcommand \nt {\noindent}

\newcommand \bvec{\left( \begin{array}{c} }
\newcommand \evec{\end{array} \right)}

\newcommand \bea{\begin{eqnarray} }
\newcommand \eea{\end{eqnarray} }
\newcommand \nn {\nonumber}
\newcommand {\be} {\begin{equation}}
\newcommand {\ee} {\end{equation}}

\newcommand {\ata} {& \times &}

\begin{document}

\title{Cherenkov Radiation from Jets in Heavy-ion Collisions}

\author{V.~Koch}
\author{A.~Majumder}
\author{Xin-Nian Wang}
\affiliation{Nuclear Science Division,
Lawrence Berkeley National Laboratory,
1 Cyclotron road, Berkeley, CA 94720}

\date{ \today}

\begin{abstract} 
  The possibility of Cherenkov-like gluon bremsstrahlung in dense matter is
  studied. We point out that the occurrence of Cherenkov radiation in dense
  matter is sensitive to the presence of partonic bound states. This 
  is illustrated by a calculation of the dispersion relation
  of a massless particle in a simple model in which it couples to two 
  different massive resonance states. We further argue that detailed 
  spectroscopy of jet correlations can directly probe the index of 
  refraction of this matter, which in turn will provide information about 
  the mass scale of these partonic bound states.

\end{abstract}


\preprint{LBNL-58447}

\maketitle
The goal of high-energy heavy-ion collisions is to create and explore 
a novel state of matter in which quarks and gluons are deconfined over 
distances considerably larger than that of a hadron. Lattice QCD 
calculations \cite{Karsch:2003jg} have predicted such a transition
with a rapid rise in the entropy density at a critical temperature 
of about $T_c \simeq 170~{\rm MeV}$. However, the entropy density 
is seen to level off somewhat below the ideal gas limit. Calculations 
with a more sophisticated resummation of quasi-particle modes \cite{Andersen:1999va} 
within the hard-thermal-loop approximation \cite{htl} improve upon the 
ideal gas picture but are still above the lattice QCD results near and 
just above $T_c$, suggesting that the plasma may possess a somewhat 
more complex structure in this regime. Indeed, recent lattice QCD 
calculations of spectral functions \cite{karsch2,hatsuda} find the
presence of charmonium states above $T_c$. This has led to the suggestion 
that at moderate temperatures, $T \simeq 1-2 \,T_c$, there exist
many bound states \cite{Shuryak:2004tx} both in the color singlet
and other colored representations, though lattice QCD results on 
baryon-strangeness correlations can rule out the 
presence of many light bound states involving only quarks and 
anti-quarks \cite{Koch:2005vg}. Furthermore, strong collective flow 
observed in experiments at the Relativistic Heavy-Ion 
Collider (RHIC) \cite{Ackermann:2000tr} also suggest a strongly
interacting plasma. Therefore, the nature of the relevant degrees 
of freedom in the matter created at RHIC needs to be further explored.

It is the purpose of this Letter to argue and demonstrate that one can
probe the resonance structure of the dense matter via the production of 
Cherenkov-like 
soft hadrons along the path of quenched jets. 
Jet quenching or medium modification 
of the jet structure has emerged as a new diagnostic tool for the 
study of partonic properties of the dense matter \cite{Gyulassy:2003mc}. 
The modification goes beyond a mere suppression of inclusive spectra of 
leading hadrons \cite{phenix-r} and has been extended to include the 
modification of two-hadron correlations \cite{star-jet,amxnw}. 
Of particular interest for the present work is the experimental
observation that soft hadrons correlated with a quenched jet have an 
angular distribution that is peaked at a finite angle away from the 
jet \cite{star-cone,phenix-cone}, whereas they peak along 
the jet direction in vacuum.
In the picture of normal gluon bremsstrahlung induced by multiple
parton scattering, one can identify these associated soft hadrons
with those from the hadronization of radiated gluons. Because of
the Landau-Pomeranchuck-Midgal (LPM) interference, the angular
distribution of the induced gluon bremsstrahlung does peak
at an angle $\theta\sim \sqrt{2/\omega_gL}$ away from the initial
jet direction \cite{vitev,mw05}. However, this angle decreases with
the length of the jet propagation or with the centrality of the
nuclear collisions. This is currently not supported by the experimental
data \cite{star-cone,phenix-cone}.

Two other known phenomena can, however, result in such an emission 
pattern:
Mach cones generated by the
hydrodynamical propagation of energy deposited by a quenched jet
along its path \cite{Shuryak,Stoecker} and Cherenkov gluon radiation.
The angle of particle emission from the generated Mach cone is
determined by the velocity of sound which can be calculated in lattice QCD. 
In the case of Cherenkov gluon radiation, the situation is less clear. 
While the general phenomenon has been discussed \cite{Mueller,Dremin}, the 
essential question, whether the index of refraction, which determines the
cone angle, is larger than unity 
in deconfined QCD matter has not been addressed. Indeed, calculations 
in the Hard Thermal Loop (HTL) approximation of QCD  \cite{htl}
do not allow for Cherenkov gluon radiation \cite{Mueller}. Quenched 
lattice QCD calculations also indicate a time-like dispersion relation for 
large momenta ($p>>T$) at $T>T_c$ \cite{Petreczky:2001yp}. The situation,
however, is unclear for soft modes $p<T$ and at around $T_c$.

In this Letter, we start with the realization that a large index 
of refraction and therefore Cherenkov-like gluon bremsstrahlung can
only result from coherent gluon scattering off partonic bound states 
in the QGP.
The situation is analogous to photons in a
gas, where the coherent scattering off  atoms in the gas allows for
Cherenkov radiation, but not in a gas of single 
elementary charged particles. Thus the observation of Cherenkov-like
bremsstrahlung in heavy-ion collisions would serve as an signal for the 
presence of bound states in the QGP.

Obviously, a large index of refraction, corresponding to a space-like 
dispersion relation, requires attractive interaction. This is where the bound
states enter the picture: It is natural to assume that these bound states have
excitations, just like an atom, and gluon interaction can cause transition
between these bound states through simple resonant scattering.
If the energy of the gluon is smaller than that of the first excited state, 
the scattering amplitude is attractive leading to an attractive optical 
potential for such a gluon. As a consequence, the gluon dispersion relation
in this regime becomes space-like and Cherenkov radiation will occur. 
Similar effects have been noted in the context of gluon scattering
in nuclei \cite{Dremin} or pion scattering in nuclear matter, where the 
transition of the nucleon to the Delta resonance provided the necessary 
attraction \cite{bertsch_koch}.

\begin{figure}[htbp]
\resizebox{2in}{1.5in}{\includegraphics[1in,0in][6in,4in]{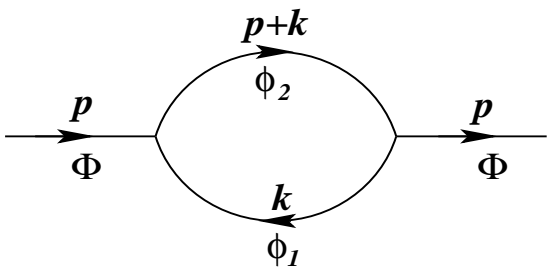}}
\caption{ The general contribution to the self-energy 
of $\Phi$ due to transitional interaction. }
\label{fig1}
\end{figure}

To illustrate this effect within finite temperature field theory, 
a simplified model is employed: a massless scalar $\Phi$ (representing 
the radiated gluon) coupled to two massive scalars $\phi_1,\phi_2$, 
representing a bound state and its excitation. The interaction is such 
that $\Phi$ may induce a transition from $\phi_1$ to $\phi_2$ and vice-versa.
The Lagrangian for such a system, ignoring the self-interactions between 
the scalars, has the form 

\bea
\mathcal{L} = \sum_{i=1}^2 \hf (\prt \phi_i)^2 + \hf (\prt \Phi)^2
+ \sum_{i=1}^2 \hf m_i^2 \phi_i^2 +  
g \Phi \phi_2 \phi_3.
\eea

The coupling constant $g$ is dimensionful; this along with all other 
scales in this Letter will be expressed in units of the temperature. 
Ignoring issues related to vacuum renormalizability of such a theory, we 
focus on a study of the dispersion relation of the massless scalar in such an 
environment. The thermal propagator of $\Phi$ in the interacting 
theory is given in general as 

\bea
D(p^0,\vp) =  \frac{1}{(p^0)^2 - |\vp|^2 - \Pi(p^0,\vp,T)}, 
\eea

\nt
and the dispersion relation is given by the on-shell condition: 
$(p^0)^2 - |\vp|^2 - \Pi(p^0,\vp,T)=0$. Here, $\Pi(p^0,\vp,T)$ is the 
thermal self-energy of $\Phi$ due to loop diagrams such as the one 
shown in Fig.\ref{fig1}. The imaginary parts of this self-energy at finite 
temperature has been discussed in Ref.~\cite{Weldon:1983jn}. In this Letter, the 
focus will be on the real part of the one-loop self energy. 

In order to discuss the the essential contributions to the self-energy, it 
is decomposed, following Ref.~\cite{Wong:2000hq}, as:

\bea 
\Pi(p^0,p) &=& g^2 \int \frac{d^3k}{(2\pi)^3}  \hf  \\ \label{phi_se}
\ata \Bigg[ \frac{1}{2E_1(\vk)} \left\{ \frac{ \{1 + n[E_1(\vk)]\} + \{n[E_1(\vk)]\} }
{[p^0 + E_1(\vk)]^2 - [E_2(\vp+\vk)]^2 } \right\} \nn \\
&+& \frac{1}{2E_1(\vk)} \left\{ \frac{ \{ 1 + n[E_1(\vk)] \}  + \{ n[E_1(\vk)]\} } 
{ [p^0 - E_1(\vk)]^2 - [E_2(\vp-\vk)]^2   }  \right\}  \nn \\
&+& \frac{1}{2E_2(\vk)} \left\{ \frac{ \{ 1 + n[E_2(\vk)]  \} + \{ n[E_2(\vk)]\} }
{[p^0 + E_2(\vk)]^2 - [E_1(\vp+\vk)]^2}  \right\}  \nn \\
&+& \frac{1}{ 2E_2(\vk) } \left\{ \frac{ \{ 1 + n[E_2(\vk)]\}  +  \{n[E_2(\vk)]\} }
{[p^0 - E_2(\vk)]^2 - [E_1(\vp - \vk)]^2 }  \right\} \Bigg], \nn
\eea

\begin{figure}[htbp]
\resizebox{3in}{3in}{\includegraphics[0in,0in][6in,6in]{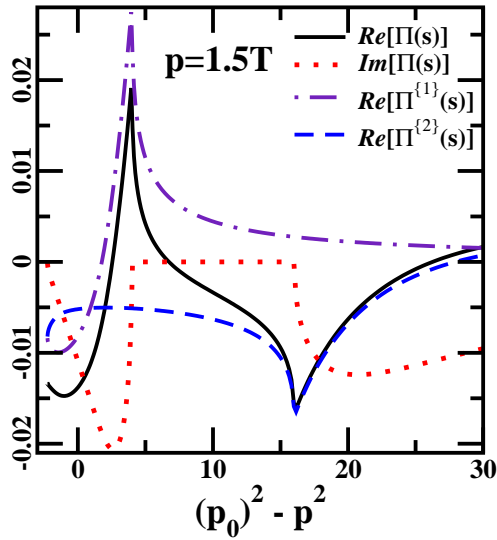}}
    \caption{The real (solid) and  imaginary part (dotted) of the full 
      self-energy and the first two contributions (dashed and dot-dashed)
      from Eq.~(\ref{phi_se}).}
    \label{fig2}
\end{figure}

\nt
where, $n(E)$ denotes the thermal (Bose-Einstein) distribution. 
In the subsequent discussion, only the temperature dependent parts 
of the self-energy, {\it i.e.} those terms involving thermal distributions, 
are relevant. In the case when $m_2 - m_1 \gg T$, the last two terms are 
suppressed by Boltzmann factors as compared to the first two. 
 In the above equation, the first term represents the
  standard resonant scattering contribution, where the ``gluon'', $\Phi$,
  absorbs the lower bound state, $\phi_1$, and propagates
  through the space-like off-shell intermediate state $\phi_2$. 
  At low gluon momenta $p$, this is the dominant term which provides 
  the necessary attraction. In Fig.\ref{fig2}, the real and 
  imaginary parts of the self-energy are plotted as functions of
  the energy for a fixed momentum $|\vec{p}| = 1.5 T$, $m_1=T$, $m_2=3T$ 
  and $g=2T$. Contributions to the real part from the first two terms
  which are  the dominant contributions are also plotted. Note that at
  low energies, below the resonance ($s\equiv p_0^2-\vec{p}^2=(m_2-m_1)^2$),
  we find attraction as expected from resonance scattering. At higher 
  energies, just before the threshold ($s=(m_1+m_2)^2$) for the
  production of a pair of $\phi_1,\phi_2$ states, additional attraction 
  is also found for the given four-momentum. This corresponds to the 
  emission of a $\phi_1$ by $\Phi$ and creation of an space-like off-shell $\phi_2$. 
  This region, however, is not relevant for the subsequent discussion of
  Cherenkov radiation.

In principle, one should include the contribution from the
self-coupling of $\Phi$ to the self energy. For vector gluons, such
a contribution, in isolation, gives rise to a time-like dispersion relation, 
and in combination with the gluon-resonance coupling may lead to a 
complicated momentum dependence of the dispersion relation.
However, if the 
gluon-resonance coupling is much stronger than the gluon self-coupling,
one may neglect such a contribution to the self-energy.

\begin{figure}[htbp]
\resizebox{2.3in}{2.3in}{\includegraphics[1.5in,0.75in][6.5in,5.75in]{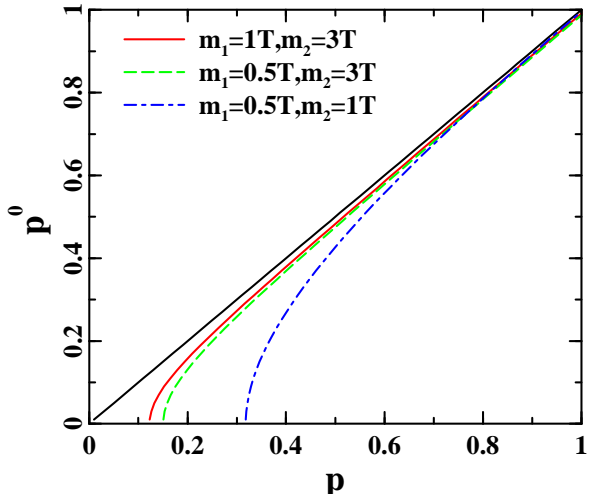}}
    \caption{ The dispersion relation of $\Phi$ in a thermal medium
      with transitional coupling to two massive particles. The 
      diagonal line represents the light-cone.}
    \label{fig3}
\end{figure}

The resulting dispersion relations for different choices of masses 
of $\phi_1, \phi_2$ are shown in Fig.~\ref{fig3}. As expected we 
obtain a space-like dispersion relation in low momentum which approaches 
the light-cone as $(p^0,p)$ is increased. Even though we have studied 
a simple scalar theory, the attraction leading to Cherenkov-like 
bremsstrahlung has its origin in resonant scattering. Thus, the result 
is genuine and only depends on the masses of the bound
states and their excitations.

\begin{figure}[htbp]
\resizebox{2.3in}{2.3in}{\includegraphics[1.5in,0.7in][6.75in,5.75in]{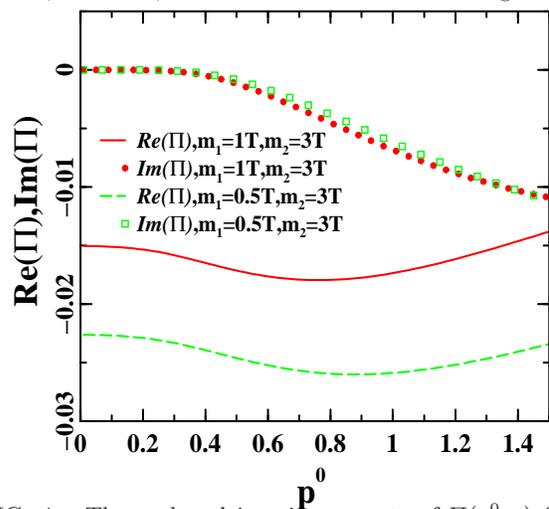}}
\caption{ The real and imaginary parts of $\Pi(p^0,p)$ for $p^0,p$ which 
satisfy the quasi-particle dispersion relation. The choice of parameters and 
the legends for the real parts are the same as in Fig.~\ref{fig3}.  }
    \label{fig4}
\end{figure}

Another essential issue is the behavior of the imaginary part of the 
self-energy along the curve of the quasi-particle dispersion relation. 
The imaginary parts are always found to be negative, which in our 
convention indicates damping of the modes. In Fig.~\ref{fig4} we plot 
the real and imaginary parts of the self-energies  for values of $(p_0,p)$
which satisfy the in-medium dispersion relation. The two sets of 
curves correspond to the first two sets of parameters in Fig.~\ref{fig3}.
One notes that the real part has only moderate variation in this range
of momentum. In contrast, the imaginary part seems to rise in magnitude. 
A large imaginary part indicates that the mode experiences strong damping 
and will not propagate far in the medium. However, there seems to exist 
a range of soft energies and momenta in the dispersion relation 
where the imaginary part is very small allowing the possibility for 
long range propagation of Cherenkov-like gluons. Even in the region
where the imaginary part is large and the mode is considerably damped, there
could still be a unique angular distribution of Cherenkov-like
gluon bremsstrahlung \cite{mw05} except that the energy of these
gluons is absorbed by the medium. The propagation of this energy 
through the medium can also cause sonic shock waves, however, with 
modified Mach cone angles.

\begin{figure}[htbp]
\begin{center}
\resizebox{2.3in}{2.3in}{\includegraphics[1.5in,1in][6.5in,6in]{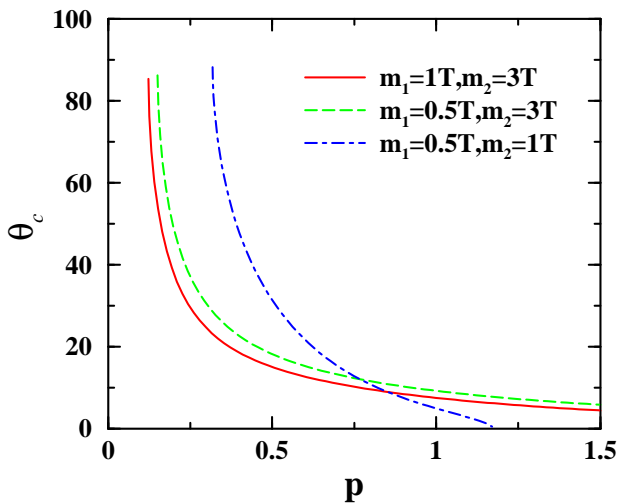}}
\end{center}
    \caption{Dependence of the Cherenkov angle on momentum of the emitted 
      particle.}
    \label{fig5}
\end{figure}

Cherenkov-like gluon bremsstrahlung may be observed as conical 
structures in two-particle correlations in jets. Shown in Fig.~\ref{fig5}
is the dependence of the Cherenkov angle $\cos\theta_c=1/n(p)$ on 
the gluon momentum, as determined from the dispersion relation 
in Fig.~\ref{fig3}. It has a strong momentum dependence and 
vanishes quickly at large gluon momentum as the dispersion relation
approaches the light-cone. Such a momentum dependence of the
emission angle is in contradistinction with that of a Mach cone, which
is independent of the momentum of the emitted particle.

The normal Cherenkov radiation (without multiple scattering of the
propagating energetic parton) also contributes to the parton energy loss.
Adopting the results obtained for photon Cherenkov radiation \cite{jackson} 
we can estimate this energy loss by
\bea
\frac{dE}{dx} = 4 \pi \alpha_s \int_{n(p_0)>1} p_0 \left[1 -
\frac{1}{n^2(p_0)}\right] dp_0,
\eea
where $n(p_0)=|\vec{p}|/p_0$ is the index of reflection.
Using the dispersion relation in our simple model, the typical
energy scale for the soft mode where Cherenkov radiation can happen
is $p_0\sim T$. The Cherenkov energy loss is about $dE/dx\sim 0.1$ 
GeV/fm for $T\sim 300$ MeV.
This is much smaller than the normal radiative energy loss induced
by multiple scattering of the energetic partons \cite{Gyulassy:2003mc}.
As discussed in Ref.~\cite{mw05}, bremsstrahlung, induced by multiple 
scattering, of soft 
gluons with a space-like dispersion relation can still lead to
Cherenkov-like angular distributions due to Landau-Pomeranchuck-Migdal
interference.  Such Cherenkov-like gluon bremsstrahlung will lead to
a similar emission pattern of soft hadrons as pure Cherenkov radiation. 
Thus, a distinctive experimental signature of the Cherenkov-like gluon
radiation is the strong momentum dependence of the emission angle of
soft hadrons leading to the disappearance of a cone-like structure
for large $p_T$ asssociated hadrons.

In conclusion, we have shown how bound states in the QGP or more generally
additional mass scales give rise to radiation of Cherenkov gluons off a
fast jet traversing the medium. These Cherenkov gluons lead to a cone-like
emission pattern of soft hadrons. The cone angle with respect to the jet
direction exhibits a strong momentum dependence in contrast to a
Mach-cone. Pure Cherenkov radiation leads to energy loss which, however, 
is too small to account for the jet suppression observed in RHIC
experiments. Collision-induced Cherenkov-like bremsstrahlung \cite{mw05} 
can explain both the observed energy loss and the emission pattern of 
soft hadrons in the direction of quenched jets.

We thank F. Karsch and E. Shuryak for helpful discussions.
This work was supported by the Director, Office of Energy
Research, Office of High Energy and Nuclear Physics, Divisions of 
Nuclear Physics, of the U.S. Department of Energy under Contract 
No. DE-AC02-05CH11231.

\end{document}